\documentclass{pj}
\usepackage{graphicx}

\begin{document}
\setcounter{page}{1}
\pjheader{Vol.\ x, y--z, 2014}

\title[]
{Symplectic Pseudospectral Time-Domain Scheme for Solving Time-Dependent Schr\"odinger Equation}
 \footnote{\it Received date}  \footnote{\hskip-0.12in*\, Corresponding
author:Jing~Shen(jingsh@hfnu.edu.cn).}
\footnote{\hskip-0.12in\textsuperscript{1} School of Electronics and Information Engineering, Hefei Normal University, Hefei, Anhui, China. \textsuperscript{2} School of Electronics and Information Engineering, Anhui University, Hefei, Anhui, China.
\textsuperscript{3} School of Electronics and Information Engineering, Zhejiang University, Hangzhou, Zhejiang, China.}

\author{Jing~Shen\textsuperscript{*,1}, Wei E.I.~Sha\textsuperscript{3}, Xiaojing~Kuang\textsuperscript{1}, Jinhua~Hu\textsuperscript{1}, Zhixiang~Huang\textsuperscript{2}, Xianliang~Wu\textsuperscript{2}}

\runningauthor{Shen, Sha, et. al.}

\tocauthor{Shen, Sha, et. al.}

\begin{abstract}
A symplectic pseudospectral time-domain (SPSTD) scheme is developed to solve Schr\"{o}dinger equation. Instead of spatial finite differences in conventional finite-difference time-domain (FDTD) methods, the fast Fourier transform is used to calculate the spatial derivatives. In time domain, the scheme adopts high-order symplectic integrators to simulate time evolution of Schr\"{o}dinger equation. A detailed numerical study on the eigenvalue problems of 1D quantum well and 3D harmonic oscillator is carried out. The simulation results strongly confirm the advantages of the SPSTD scheme over the traditional PSTD method and FDTD approach. Furthermore, by comparing to the traditional PSTD method and
the non-symplectic Runge-Kutta (RK) method, the explicit SPSTD scheme which is an infinite order of accuracy in space domain and energy-conserving in time domain, is well suited for a long-term simulation.\end{abstract}


\setlength {\abovedisplayskip} {6pt plus 3.0pt minus 4.0pt}
\setlength {\belowdisplayskip} {6pt plus 3.0pt minus 4.0pt}

\

\section{Introduction}
\label{section label}

Numerical solution to Schr\"odinger equation has become increasingly
important because of the tremendous demands for the design and optimization
of nanodevices, where quantum effects are significant or dominate $^{[1]}$. The
eigenvalue problem of Schr\"{o}dinger equation is fundamentally important for
quantum transport and nanodevice modeling. One of commonly adopted methods
to solve the eigenvalue problem of Schr\"{o}dinger equation is
FDTD method $^{[2,3]}$. In the FDTD method, spatial
derivatives in Schr\"{o}dinger equation are approximated by finite differences. The
Yee algorithm has a second-order accuracy both in space and time. Thus, a fine
discretization is required to obtain a desired result tailored to physical designs. To reduce
the complexity of time-domain solutions by decreasing grid density, we employ an
efficient and accurate approach called pseudospectral method. The pseudospectral method has
an infinite order of accuracy since Fourier transform is utilized to represent the spatial derivatives $^{[4,5]}$.
Numerical experiments have shown that the pseudospectral time-domain (PSTD) method is a factor of
${4^D} - {8^D}$ more efficient than the FDTD method (where $D$ is the dimension number $^{[6-9]}$.

Many important physical phenomena can be modeled by Hamiltonian
differential equations $^{[11,12]}$. The time evolution of Hamiltonian is essentially a symplectic transform;
Equivalently, Hamiltonian flow conserves the symplectic structure $^{[11-16]}$. The symplectic schemes
are the time-steeping strategies designed to preserve the global symplectic structure of
the phase space of a Hamiltonian system. Symplectic schemes have proven themselves to
be one of best candidates for numerically modeling the Hamiltonian system,
especially for a long-term simulation. The symplectic scheme has been
applied to solve Schr\"{o}dinger equation, and numerical examples have been
shown $^{[17-18]}$. In this letter, we integrate the pseudospectral method with symplectic schemes to
construct a symplectic pseudospectral time-domain (SPSTD) scheme for solving Schr\"{o}dinger equations.

\section{Theory}
\label{sec:formulation}

\subsection{Construction of the Algorithm}
The time-dependent Schr\"{o}dinger equation is given by $^{[2]}$

\begin{equation}\label{1}
  i\hbar \frac{{\partial \psi \left( {{\bf{r}},t} \right)}}{{\partial t}} =  - \frac{{{\hbar ^2}}}{{2{m^ * }}}{\nabla ^2}\psi \left( {{\bf{r}},t} \right) + V\left( {\bf{r}} \right)\psi \left( {{\bf{r}},t} \right)
\end{equation}
where $\psi \left( {{\bf{r}},t} \right)$ is the wave function that is a probability amplitude describing the quantum state of a particle at the position ${\bf{r}}$ and time $t$, ${m^ * }$ is the (effective) mass of the particle, $- \frac{{{\hbar ^2}}}{{2{m^ * }}}{\nabla ^2}$ is the kinetic energy operator, $V\left( {\bf{r}} \right)$ is the time-independent potential energy, and $-\frac{{{\hbar ^2}}}{{2{m^ * }}}{\nabla ^2} + V$ is the Hamiltonian operator. To avoid using complex numbers, one can separate the variable $\psi \left( {{\bf{r}},t} \right)$ into its real and imaginary parts as
\begin{equation}\label{2}
\psi \left( {{\bf{r}},t} \right) = {\psi _R}\left( {{\bf{r}},t} \right) + i{\psi _I}\left( {{\bf{r}},t} \right).
\end{equation}

Inserting Eq. (2) into Eq. (1), we can get the following coupled set of equations $^{[3]}$
\begin{equation}\label{3}
\hbar \frac{{\partial {\psi _R}\left( {{\bf{r}},t} \right)}}{{\partial t}} =  - \frac{{{\hbar ^2}}}{{2{m^ * }}}\left[ \begin{array}{l}
\frac{{{\partial ^2}{\psi _I}\left( {{\bf{r}},t} \right)}}{{\partial {x^2}}}\\
 + \frac{{{\partial ^2}{\psi _I}\left( {{\bf{r}},t} \right)}}{{\partial {y^2}}}\\
 + \frac{{{\partial ^2}{\psi _I}\left( {{\bf{r}},t} \right)}}{{\partial {z^2}}}
\end{array} \right] + V\left( {\bf{r}} \right){\psi _I}\left( {{\bf{r}},t} \right),
\end{equation}

\begin{equation}\label{4}
\hbar \frac{{\partial {\psi _I}\left( {{\bf{r}},t} \right)}}{{\partial t}} = \frac{{{\hbar ^2}}}{{2{m^ * }}}\left[ \begin{array}{l}
\frac{{{\partial ^2}{\psi _R}\left( {{\bf{r}},t} \right)}}{{\partial {x^2}}}\\
 + \frac{{{\partial ^2}{\psi _R}\left( {{\bf{r}},t} \right)}}{{\partial {y^2}}}\\
 + \frac{{{\partial ^2}{\psi _R}\left( {{\bf{r}},t} \right)}}{{\partial {z^2}}}
\end{array} \right] - V\left( {\bf{r}} \right){\psi _R}\left( {{\bf{r}},t} \right).
\end{equation}

A mesh is defined in a discrete set of grid points that sample the wave
function in space and time. The real and imaginary parts of
the wave function can be represented as
\begin{equation}\label{5}
{\psi _R}\left( {{\bf{r}},t} \right) \approx \psi _R^n(i,j,k) = {\psi _R}(i{\Delta _x},j{\Delta _y},k{\Delta _z},n{\Delta _t}),
\end{equation}
\begin{equation}\label{6}
{\psi _I}\left( {{\bf{r}},t} \right) \approx \psi _I^n(i,j,k) = {\psi _I}(i{\Delta _x},j{\Delta _y},k{\Delta _z},n{\Delta _t}),
\end{equation}
where ${\Delta _x}$ , ${\Delta _y}$ , and ${\Delta _z}$  are, respectively, the spatial steps in the $x$, $y$, and $z$ coordinate directions, ${\Delta _t}$ is the time step, and $i$, $j$, $k$ and $n$ are integers.

Regarding the pseudospectral method in space, we take the Fourier series expansion
\begin{equation}\label{7}
f(x) = \sum\limits_{n =  - \infty }^{ + \infty } {{a_n}{e^{i{K_n}x}}},
\end{equation}
with
\begin{equation}\label{8}
{a_n} = \frac{1}{L}\int_0^L {f\left( x \right)} {e^{ - i{K_n}x}}{\rm{d}}x = {F_x}\left[ {f\left( x \right)} \right],
\end{equation}
where $L$ is the periodicity of the structure, ${K_n} = {{2\pi n} \mathord{\left/
 {\vphantom {{2\pi n} L}} \right.
 \kern-\nulldelimiterspace} L}$, $n = 0, \pm 1, \pm 2, \cdots$, and ${F_x}$ stands for the forward Fourier transforms in the $x$ direction. The corresponding spatial derivatives can be obtained by
\begin{equation}\label{9}
\frac{{{\rm{d}}f}}{{{\rm{d}}x}} = \sum\limits_{n =  - \infty }^{ + \infty } {i{K_n}{a_n}{e^{i{K_n}x}}}  = F_x^{ - 1}\left\{ {i{K_n}{F_x}\left[ {f\left( x \right)} \right]} \right\},
\end{equation}
where $F_x^{ - 1}$ stands for the inverse Fourier transforms in the $x$ direction. The forward and inverse Fourier transforms can be fast and efficiently computed by fast Fourier transform (FFT) algorithms. For second-order derivatives, we have
\begin{equation}\label{10}
\frac{{{\partial ^2}{\psi _I}}}{{\partial {x^2}}} = F_x^{ - 1}\left\{ {i{K_n}{F_x}\left[ {F_x^{ - 1}\left\{ {i{K_n}{F_x}\left[ {{\psi _I}\left( {{\bf{r}},t} \right)} \right]} \right\}} \right]} \right\}.
\end{equation}

It should be noted that only real parts of results are remained after each inverse Fourier transform. Therefore, with the help of Fourier transforms, Eqs. (3) and (4) can be rewritten as
\begin{equation}\label{11}
\begin{array}{l}
\psi _R^{n + 1}\left( {i,j,k} \right) = \psi _R^n\left( {i,j,k} \right)\\
 - \frac{\hbar \Delta t}{{2{m^ * }}}\left[ {{F_x^{ - 1}}\left( {i{K_x}F_x\left( {{F_x^{ - 1}}\left( {i{K_x}F_x\left( {\psi _I^n\left( {i,j,k} \right)} \right)} \right)} \right)} \right)} \right]\\
 - \frac{\hbar \Delta t}{{2{m^ * }}}\left[ {{F_y^{ - 1}}\left( {i{K_y}F_y\left( {{F_y^{ - 1}}\left( {i{K_y}F_y\left( {\psi _I^n\left( {i,j,k} \right)} \right)} \right)} \right)} \right)} \right]\\
 - \frac{\hbar \Delta t}{{2{m^ * }}}\left[ {{F_z^{ - 1}}\left( {i{K_z}F_z\left( {{F_z^{ - 1}}\left( {i{K_z}F_z\left( {\psi _I^n\left( {i,j,k} \right)} \right)} \right)} \right)} \right)} \right]\\
 + \frac{{V\left( {i,j,k} \right)\Delta t}}{\hbar } \times \psi _I^{n + {1 \mathord{\left/
 {\vphantom {1 2}} \right.
 \kern-\nulldelimiterspace} 2}}\left( {i,j,k} \right)
\end{array}
\end{equation}

\begin{equation}\label{12}
\begin{array}{l}
\psi _I^{n + 1}\left( {i,j,k} \right) = \psi _I^n\left( {i,j,k} \right)\\
{\rm{ + }}\frac{\hbar \Delta t}{{2{m^ * }}}\left[ {F_x^{ - 1}\left( {i{K_x}{F_x}\left( {F_x^{ - 1}\left( {i{K_x}{F_x}\left( {\psi _R^n\left( {i,j,k} \right)} \right)} \right)} \right)} \right)} \right]\\
{\rm{ + }}\frac{\hbar \Delta t}{{2{m^ * }}}\left[ {F_y^{ - 1}\left( {i{K_y}{F_y}\left( {F_y^{ - 1}\left( {i{K_y}{F_y}\left( {\psi _R^n\left( {i,j,k} \right)} \right)} \right)} \right)} \right)} \right]\\
{\rm{ + }}\frac{\hbar \Delta t}{{2{m^ * }}}\left[ {F_z^{ - 1}\left( {i{K_z}{F_z}\left( {F_z^{ - 1}\left( {i{K_z}{F_z}\left( {\psi _R^n\left( {i,j,k} \right)} \right)} \right)} \right)} \right)} \right]\\
 - \frac{{V\left( {i,j,k} \right)\Delta t}}{\hbar } \times \psi _R^{n + {1 \mathord{\left/
 {\vphantom {1 2}} \right.
 \kern-\nulldelimiterspace} 2}}\left( {i,j,k} \right)
\end{array}
\end{equation}

In order to implement the symplectic algorithm, we notate a wave function of space and time at a discrete stage in the time step as
\begin{equation}\label{13}
\begin{array}{l}
\psi (i,j,k) = {\psi ^{n + {l \mathord{\left/
 {\vphantom {l m}} \right.
 \kern-\nulldelimiterspace} m}}}\left( {i{\Delta _x},j{\Delta _y},k{\Delta _z},(n + {\tau _l}){\Delta _t}} \right)
 \end{array}
\end{equation}
where $n + l/m$  denotes the $lth$  stage after $n$  time steps, $m$  is the total stage number, and ${\tau _l}$  is the fixed time with respect to the $lth$  stage. With the help of Eqs. (3) and (4), the Schr\"{o}dinger equation can be casted into a matrix form
\begin{equation}\label{14}
\frac{\partial }{{\partial t}}\left( {\begin{array}{*{20}{c}}
{{\psi _R}}\\
{{\psi _I}}
\end{array}} \right) = L\left( {\begin{array}{*{20}{c}}
{{\psi _R}}\\
{{\psi _I}}
\end{array}} \right) = (A + B)\left( {\begin{array}{*{20}{c}}
{{\psi _R}}\\
{{\psi _I}}
\end{array}} \right)
\end{equation}

\begin{equation}\label{15}
A = \left( {\begin{array}{*{20}{c}}
0&K\\
0&0
\end{array}} \right),
B = \left( {\begin{array}{*{20}{c}}
0&0\\
{ - K}&0
\end{array}} \right)
\end{equation}

\begin{equation}\label{16}
K =  - \frac{\hbar }{{2{m^ * }}}\left( {\frac{\partial }{{\partial {x^2}}} + \frac{\partial }{{\partial {y^2}}} + \frac{\partial }{{\partial {z^2}}}} \right) + \frac{V}{\hbar },
\end{equation}
where ${A^v} = 0$ and ${B^v} = 0$  if $v \ge 2$. It is easy to prove that $L$ in Eq. (14) is an asymmetric operator and therefore the exact solution of Schr\"{o}dinger equation $\exp (Lt)$ is an orthogonal operator conserving the total energy of quantum system. Using the product of elementary symplectic mapping, the exact solution of (14) from $t = 0$  to $t = {\Delta _t}$  can be approximately
\begin{equation}\label{17}
\exp ({\Delta _t}(A + B)) = \prod\limits_{l = 1}^m {\exp ({d_l}{\Delta _t}B)} \exp ({c_l}{\Delta _t}A) + O({\Delta _t}^{p + 1}) = \prod\limits_{l = 1}^m {(1 + {d_l}{\Delta _t}B)(} 1 + {c_l}{\Delta _t}A) + O({\Delta _t}^{p + 1})
\end{equation}
where ${c_l}$  and  ${d_l}$ are the coefficients of symplectic integrators, and $p$  is the order of the approximation. The symplectic integrators can satisfy the time-reversible or symmetric condition $^{[19,20]}$. The detailed update equation for the real part of the wave function at the $lth$  stage can be written as

\begin{equation}\label{18}
\begin{array}{l}
\psi _R^{n+l/m}\left( {i,j,k} \right) = \psi _R^{n+(l-1)/m}\left( {i,j,k} \right)\\
 - \alpha\left[ {{F_x^{ - 1}}\left( {i{K_x}F_x\left( {{F_x^{ - 1}}\left( {i{K_x}F_x\left( {\psi _I^{n+l/m}\left( {i,j,k} \right)} \right)} \right)} \right)} \right)} \right]\\
 - \alpha\left[ {{F_y^{ - 1}}\left( {i{K_y}F_y\left( {{F_y^{ - 1}}\left( {i{K_y}F_y\left( {\psi _I^{n+l/m}\left( {i,j,k} \right)} \right)} \right)} \right)} \right)} \right]\\
 - \alpha\left[ {{F_z^{ - 1}}\left( {i{K_z}F_z\left( {{F_z^{ - 1}}\left( {i{K_z}F_z\left( {\psi _I^{n+l/m}\left( {i,j,k} \right)} \right)} \right)} \right)} \right)} \right]\\
 + \frac{{V\left( {i,j,k} \right)c_l\Delta t}}{\hbar } \times \psi _I^{n+l/m}\left( {i,j,k} \right)
\end{array},
\end{equation}
where $\alpha=\frac{\hbar c_l\Delta_t}{{2{m^ * }}}$. Here the fourth-order symmetric symplectic integrators are employed, i.e. ${c_1} = 0.26833010$, ${c_2} =  - 0.18799162$, ${c_3} = 0.91966152$, and ${d_l} = {c_{m - l + 1}}$ ($1 \le l \le m$).

\subsection{Stability Analysis of SPSTD Algorithm and Boundary Conditions}

According to the von Neumann stability method, the solution of the wave function can be represented as a superposition of plane-waves
\begin{equation}\label{19}
\psi (x,y,z,t) = {A_0}\exp \left( { - {j_0}({k_x}x + {k_y}y + {k_z}z)} \right)
\end{equation}
where${k_x} = {k_0}\sin \theta \cos \varphi,{k_y} = {k_0}\sin \theta \sin \varphi,{k_z} = {k_0}\cos \theta $,${k_0} = \frac{{{p_m}}}{\hbar }$  is the wave number, ${p_m}$ is the momentum, and $\theta$ and $\varphi$  are the spherical angles. The
$qth-order$ collocated differences are used to discretize the second-order spatial derivatives, i.e.
\begin{equation}\label{20}
\frac{{{\partial ^2}\psi }}{{\partial {z^2}}} = \frac{{{\partial ^2}{A_0}\exp \left( { - {j_0}\left( {{k_x}x + {k_y}y + {k_z}z} \right)} \right)}}{{\partial {z^2}}} =  - k_z^2\psi,
\end{equation}

For simplicity, we consider a 1D Schr\"odinger equation with zero potential energy
\begin{equation}\label{21}
\frac{\partial }{{\partial t}}\left( {\begin{array}{*{20}{c}}
{{\psi _R}}\\
{{\psi _I}}
\end{array}} \right) = \left( {\begin{array}{*{20}{c}}
0&{ - \frac{\hbar }{{2{m^ * }}}\frac{{{\partial ^2}}}{{\partial {z^2}}}}\\
{\frac{\hbar }{{2{m^ * }}}\frac{{{\partial ^2}}}{{\partial {z^2}}}}&0
\end{array}} \right)\left( {\begin{array}{*{20}{c}}
{{\psi _R}}\\
{{\psi _I}}
\end{array}} \right),
\end{equation}
and corresponding spatial discretization form is given by
\begin{equation}\label{22}
\frac{\partial }{{\partial t}}\left( {\begin{array}{*{20}{c}}
{{\psi _R}}\\
{{\psi _I}}
\end{array}} \right) = \left( {\begin{array}{*{20}{c}}
0&{\frac{\hbar }{{2{m^ * }}}{k_z}}\\
{ - \frac{\hbar }{{2{m^ * }}}{k_z}}&0
\end{array}} \right)\left( {\begin{array}{*{20}{c}}
{{\psi _R}}\\
{{\psi _I}}
\end{array}} \right),
\end{equation}

It is trivial to access the discretized evolution matrix ${L^d}$ with the high-order symplectic integration scheme
\begin{equation}\label{23}
{L^d} = \left[ {\begin{array}{*{20}{c}}
{{l_{11}}}&{{l_{12}}}\\
{{l_{21}}}&{{l_{22}}}
\end{array}} \right] = \prod\limits_{l = 1}^m {\left( {\begin{array}{*{20}{c}}
1&0\\
{ - \frac{\hbar }{{2{m^ * }}}{k_z}{d_l}{\Delta _t}}&1
\end{array}} \right)\left( {\begin{array}{*{20}{c}}
1&{\frac{\hbar }{{2{m^ * }}}{k_z}{c_l}{\Delta _t}}\\
0&1
\end{array}} \right)},
\end{equation}

The eigenvalues $\lambda$   of the evolution matrix satisfy the following eigen-equation
\begin{equation}\label{24}
{\lambda ^2} - tr({L^d})\lambda  + \det ({L^d}) = 0 ,
\end{equation}
where $tr({L^d})$ and $\det ({L^d})$  are the trace and determinant of the evolution matrix, respectively. Regarding that the discretized evolution matrix is a symplectic matrix with the determinant of 1. The eigen-equation then can be simplified as
\begin{equation}\label{25}
{\lambda ^2} - tr({L^d})\lambda  + 1 = 0,
\end{equation}
and its solutions are ${\lambda _{1,2}} = \frac{{tr({L^d}) \pm {j_0}\sqrt {4 - {{\left[ {tr({L^d})} \right]}^2}} }}{2}$ . A stable algorithm requires $|{\lambda _{1,2}}| = 1$ , and thus $|tr({L^d})| \le 2$ . Implementing terms of matrix multiplications, we can get
\begin{equation}\label{26}
tr({L^d}) = 2 + \sum\limits_{l = 1}^m {{{( - 1)}^l}{g_l}} {\left( {{{\left( {\frac{\hbar }{{2{m^ * }}}} \right)}^2}\Delta _t^2k_z^2} \right)^l}
\end{equation}

\begin{equation}\label{27}
tr({L^d}) = 2 + \sum\limits_{l = 1}^m {{{( - 1)}^l}{g_l}} {\left( {{{\left( {\frac{\hbar }{{2{m^ * }}}} \right)}^2}\Delta _t^2k_z^2} \right)^l}
\end{equation}

\begin{equation}\label{28}
{g_l} = \sum\limits_{1 \le {i_1} \le {j_1} < {i_2} \le {j_2} <  \cdots  < {i_l} \le {j_l} \le m} {{c_{{i_1}}}{d_{{j_1}}}{c_{{i_2}}}{d_{{j_2}}} \cdots } {c_{{i_l}}}{d_{{j_l}}} + \sum\limits_{1 \le {i_1} < {j_1} \le {i_2} < {j_2} \le  \cdots  \le {i_l} < {j_l} \le m} {{d_{{i_1}}}{c_{{j_1}}}{d_{{i_2}}}{c_{{j_2}}} \cdots } {d_{{i_l}}}{c_{{j_l}}}
\end{equation}

The above results can be generalized to a 3D Schr\"odinger equation with zero potential energy, i.e.
\begin{equation}\label{29}
tr({L^d}) = 2 + \sum\limits_{l = 1}^m {{{( - 1)}^l}{g_l}} {\left( {{{\left( {\frac{\hbar }{{2{m^ * }}}} \right)}^2}\Delta _t^2{{({k_x} + {k_y} + {k_z})}^2}} \right)^l}
\end{equation}

Finally we can get
\begin{equation}\label{30}
\sqrt {\frac{\hbar }{{{m^ * }}}\frac{{{\Delta _t}}}{{{\Delta _\delta }^2}}}  \le CFL ,
\end{equation}
where $CFL$  is the Courant-Friedrichs-Levy (CFL) number. Table 1 lists the maximum stability (CFL number) of the traditional FDTD method, PSTD approach, and SPSTD scheme. The symmetric symplectic integrators for the SPSTD scheme is given as follows:${c_1} = 0.26833010$, ${c_2} =  - 0.18799162$, ${c_3} = 0.91966152$, and ${d_l} = {c_{m - l + 1}}$($1 \le l \le m$).From the table, the stability of the SPSTD scheme$^{[6]}$ is larger than that of the traditional PSTD method through a careful optimization of symplectic integrators.
\vskip 3mm

\noindent{\footnotesize \bf Table 1. \rm The numerical stability for various algorithms. $D = 1,2,3$ is the dimension number.

\vskip 3mm \tabcolsep 8pt

\centerline{\footnotesize
\begin{tabular}{cccccc}\hline
 Algorithm & CFL Number\\
\hline
FDTD
 & $\frac{1}{{\sqrt D  }}$ \\
PSTD
 & $\frac{2}{{\sqrt D \pi }}$ \\
SPSTD
 & $1.503 \times \frac{2}{{\sqrt D \pi }}$ \\
\hline
\end{tabular}}}

\vskip 0.5\baselineskip
\vskip 3mm

To guarantee the numerical accuracy of simulation, boundary conditions should be handled properly. Regarding periodic boundary condition or fast decayed wave function, we employ discrete Fourier transform (DFT) to represent the spatial derivatives as shown in Eqs. (9). Regarding the Dirichlet boundary condition (for modeling the infinite potential well), discrete Sine Transform (DST) should be chosen to replace the DFT. Regarding the Neumann boundary condition, discrete Cosine transform (DCT) should be adopted.

\section{Numerical Results}
\label{sec:NumericalResults}

\subsection{1D Schr\"odinger equation}
For the first example, we consider a particle in a one-dimensional (1D) infinite potential well. Regarding the simulation domain and cell size, they depend on the length of the box to be simulated and the highest eigenenergy of the particle of interest, respectively. Without the loss of generality, we choose the domain to be $L{\rm{ = }}1$ nm, the cell size $\Delta x = 0.1$ nm, the time step $\Delta t = \left( {{{{m^ * }} \mathord{\left/
 {\vphantom {{{m^ * }} {4\hbar }}} \right.
 \kern-\nulldelimiterspace} {4\hbar }}} \right)\left(\Delta {x}\right)^2{\rm{ = 0}}{\rm{.0216 }}$ fs and the iteration step ${N_{\max }} = 1024$.The eigenenergies of the quantum well are quantized as,
 \begin{equation}\label{22}
 {E_n} = \frac{{{\hbar ^2}{\pi ^2}}}{{2{m^ * }{a^2}}}{n^2},n = 1,2,3,..
 \end{equation}

In order to excite all possible modes, the delta source is located at the center of the box with two grids offset.Table 2 lists the calculated eigenfrequencies. Compared with the analytical solutions, SPSTD scheme can achieve best accuracy. Using $\Delta x = 0.05$ nm, Fig. 1 and Fig. 2 show the eigenstates corresponding to the eigenfrequencies $\omega_4$ and $\omega_5$, respectively. Both the SPSTD and PSTD schemes can achieve much better results than the FDTD method.

\vskip 3mm
\noindent{\footnotesize \bf Table 2. \rm The eigenfrequency comparisons for a 1D quantum well
\vskip 3mm \tabcolsep 8pt

\centerline{\footnotesize
\begin{tabular}{cccccc}\hline
 Algorithm & FDTD & PSTD & SPSTD & Analytical\\
\hline ${\omega _{1}}$ &0.5683	&0.5683 &0.5683	&0.5713 \\
${\omega _{2}}$ &2.2731	&2.2731	&2.2731	&2.2852 \\
${\omega _{3}}$ &4.8303	&5.1145 &5.1145 &5.1416 \\
${\omega _{4}}$ &7.9558	&9.0924	&9.0924	&9.1406 \\
${\omega _{5}}$ &11.6496 &14.4910 &14.2068 &14.2823 \\
${\omega _{6}}$ &15.3434 &20.7420 &20.4578 &20.5665 \\
${\omega _{7}}$ &18.4689 &28.1295 &27.8454 &27.9932 \\
${\omega _{8}}$ &22.7309 &37.5060 &36.6536 &36.5626 \\
\hline
\end{tabular}}}

\begin{figure}[htp]
\centering
  \includegraphics[width=14cm]{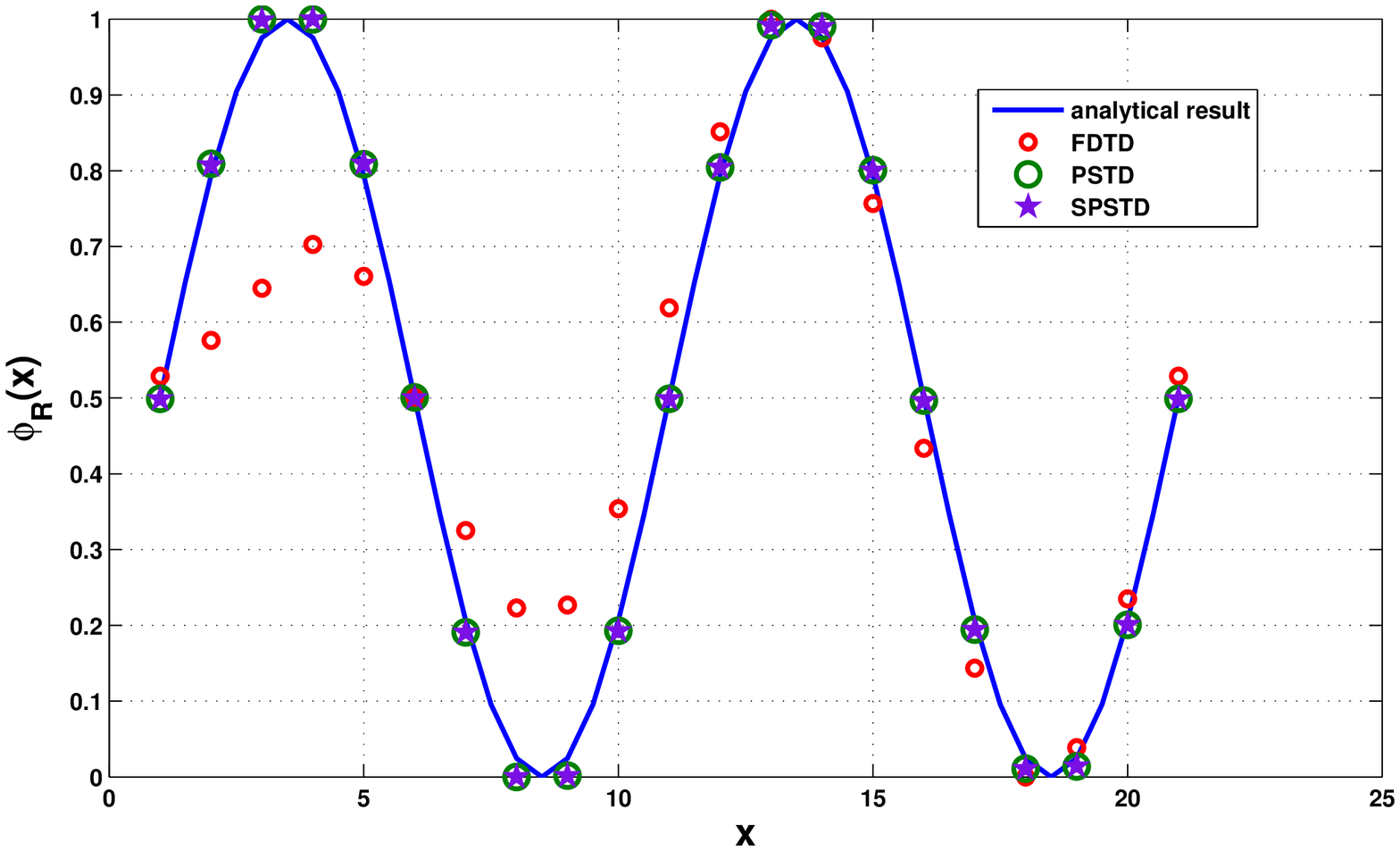}
    \caption{The normalized eigenstate (the real part of the wave function) corresponding to the eigenfrequency $\omega_4$ for a 1D quantum well.}\label{fig1.eps}
\end{figure}

\begin{figure}[htp]
\centering
  \includegraphics[width=14cm]{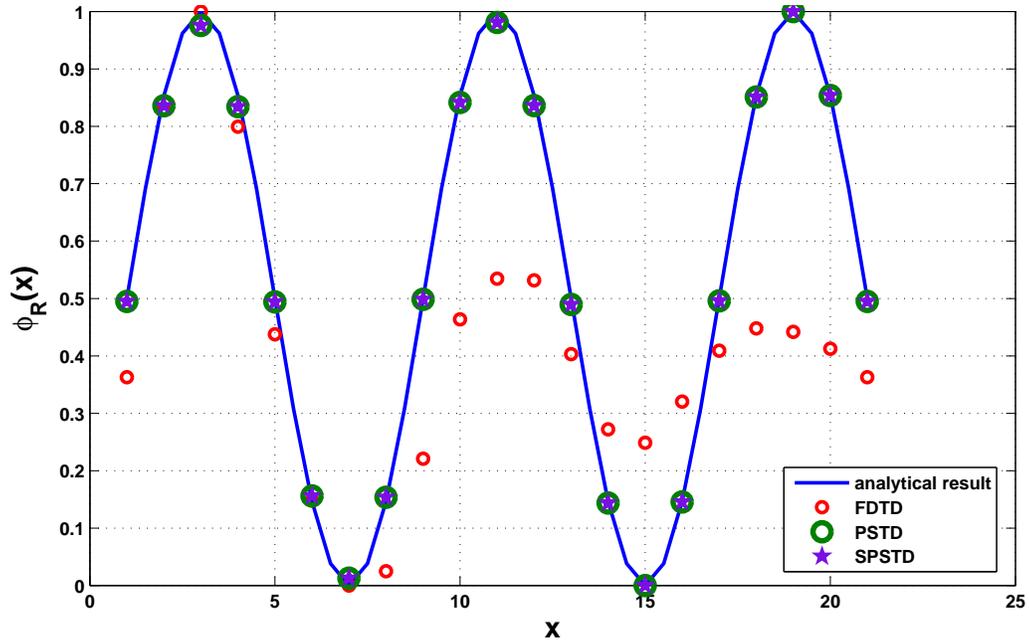}
    \caption{The normalized eigenstate (the real part of the wave function) corresponding to the eigenfrequency $\omega_5$ for a 1D quantum well.}\label{fig2.eps}
\end{figure}

The normalized condition of the wave function should be conserved under a long-term simulation, which determines energy-conserving property of Schr\"{o}dinger equation. In order to testify the property, we proceed to solve the 1D quantum well numerically using the SPSTD method, PSTD method and a non-symplectic Runge-Kutta (RK) method. In order to testify the energy-conserving property, the time evolution of the system is executed from $t = 0$ to $t =3300$ using different time steps of $\Delta t = 0.2\Delta {t_q}$ and $\Delta t = \Delta {t_q}$ ($\Delta {t_q}$ =0.0013 fs, $L=1$ nm, and $\Delta x=0.025$ nm). Fig. 3 shows the integrated wave function $\int\left|\psi(x)\right|^2 dx$ over the quantum well region by using various approaches. The SPSTD scheme holds the normalized condition of the wave function better.


\begin{figure}[htp]
\centering
  \includegraphics[width=16cm]{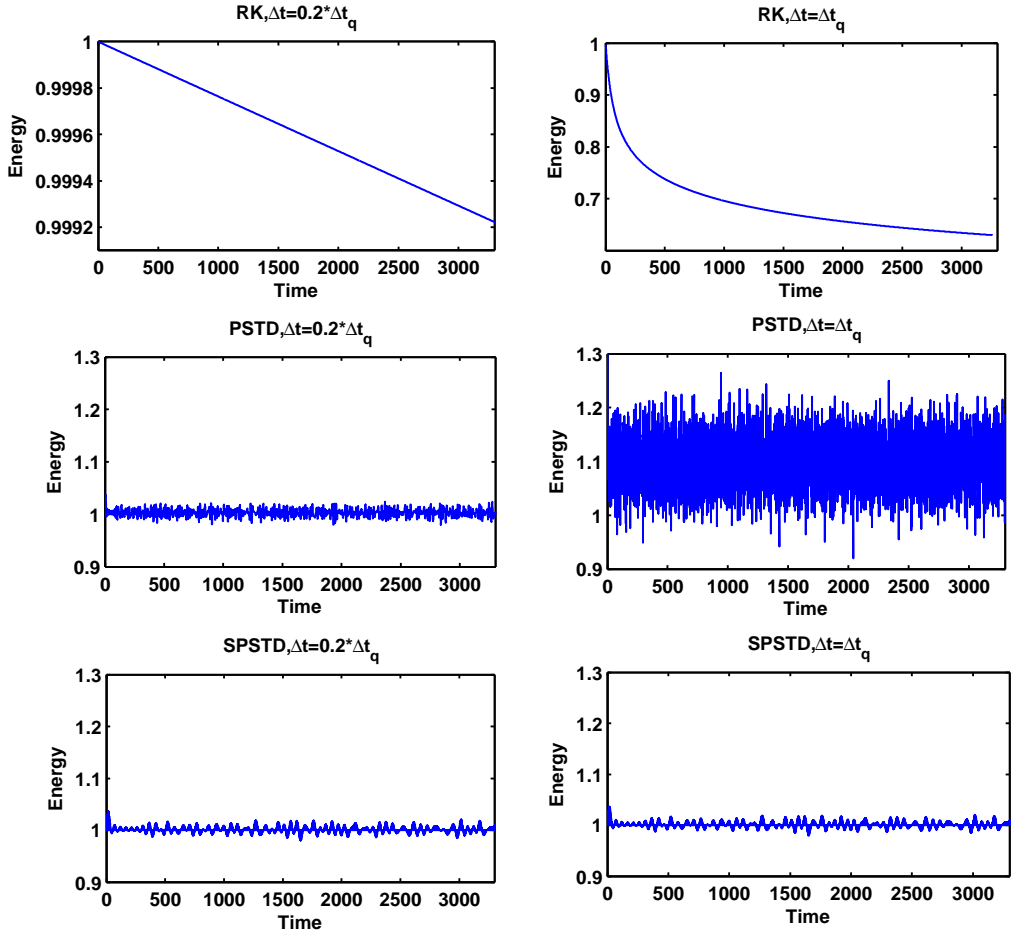}
    \caption{The time evolution of the integrated wave function $\int\left|\psi(x)\right|^2 dx$ over a 1D quantum well region.}\label{fig3.eps}
\end{figure}
\subsection{2D Schr\"odinger equation}

The simulation domain is set to ${L_x} \times {L_y} = 1\,\mathrm{nm} \times 1\,\mathrm{nm}$, $\Delta x = \Delta y  = 0.1\,\mathrm{nm}$, the time step $\Delta t = \left( {{{{m^ * }} \mathord{\left/
 {\vphantom {{{m^ * }} {8\hbar }}} \right.
 \kern-\nulldelimiterspace} {8\hbar }}} \right)\left(\Delta {x}\right)^2{\rm{ = 0}}{\rm{.0108 }}\,\mathrm{fs}$ and the iteration step ${N_{\max }} = 2048$. Table 3 lists the calculated eigenfrequencies, and Fig. 1 show the eigenstates corresponding to the eigenfrequencies $\omega_{22}$.Compared with the analytical solution, the SPSTD scheme achieves best accuracy.
\vskip 3mm
\noindent{\footnotesize \bf Table 3. \rm The eigenfrequency comparisons for a 2D quantum well
\vskip 3mm \tabcolsep 8pt
\centerline{\footnotesize
\begin{tabular}{cccccc}\hline
 Algorithm & FDTD & PSTD & SPSTD & Analytical\\
\hline ${\omega _{11}}$ &1.1366	&1.1366 &1.1366	& 1.1426\\
${\omega _{12}}$ &2.6283 &2.8413 &2.8413	&2.8565 \\
${\omega _{22}}$ & 4.1909 &4.4039 &4.5459 &4.5703 \\
${\omega _{13}}$ &5.1855 & 5.8245 &5.6825	&5.7129 \\
${\omega _{23}}$ &6.3221 &7.6718 &7.4588 &7.4268\\
${\omega _{14}}$ &8.0978 &10.0160 &9.7314 &9.7119\\
${\omega _{44}}$ &14.9169 &18.6818 & 18.2558 &18.2813 \\
\hline
\end{tabular}}}
\vskip 0.5\baselineskip
\vskip 3mm

\begin{figure}[htp]
\centering
  \includegraphics[width=16cm]{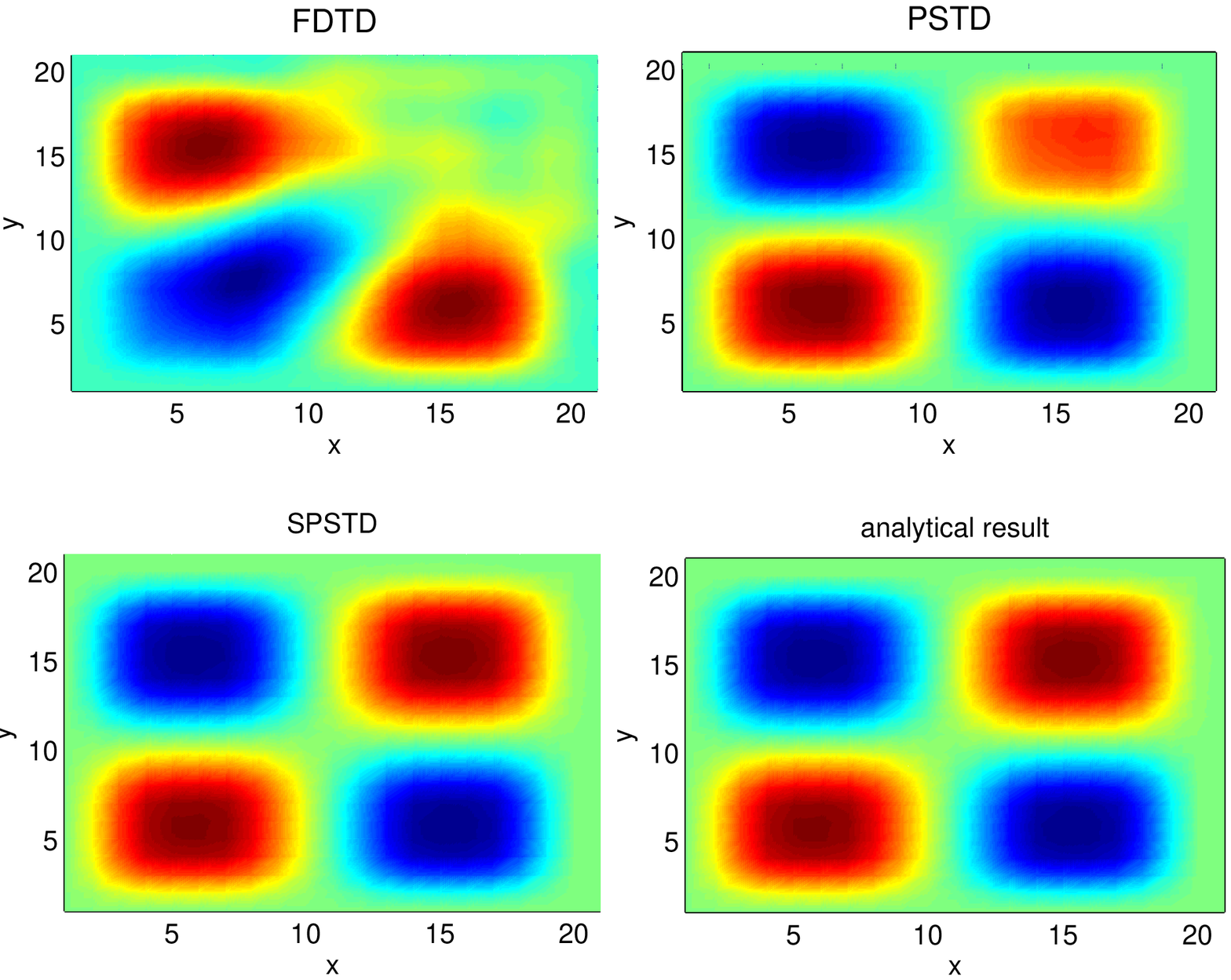}
    \caption{The normalized eigenstate (the real part of the wave function) corresponding to the eigenfrequency $\omega_{22}$ for a 2D quantum well.}\label{fig4.eps}
\end{figure}

\subsection{3D Schr\"odinger equation}

We consider a three-dimensional (3D) isotropic quantum harmonic oscillator, where the potential
energy $V(x,y,z) = \frac{1}{2}k({x^2} + {y^2} + {z^2})$. The simulation domain is set to ${L_x} \times {L_y} \times {L_z} = 1\,\mathrm{nm} \times 1\,\mathrm{nm} \times 1\,\mathrm{nm}$, $\Delta x = \Delta y = \Delta z = 0.1\,\mathrm{nm}$, the time step $\Delta t = \left( {{{{m^ * }} \mathord{\left/
 {\vphantom {{{m^ * }} {8\hbar }}} \right.
 \kern-\nulldelimiterspace} {8\hbar }}} \right)\left(\Delta {x}\right)^2{\rm{ = 0}}{\rm{.0108 }}\,\mathrm{fs}$ and the iteration step ${N_{\max }} = 2048$. The eigenenergies of the harmonic oscillator are
 \begin{equation}\label{24}
 {E_{{n_{x,}}{n_y},{n_z}}} = \left( {{n_x} + {n_y}{\rm{ + }}{n_{\rm{z}}} + 1.5} \right)\hbar \omega
 \end{equation}
 Table 4 lists the calculated eigenfrequencies. Compared with the analytical solution, the SPSTD scheme achieves best accuracy.
\\
\vskip 2mm
\noindent{\footnotesize \bf Table 4. \rm The eigenfrequency comparisons for a 3D quantum harmonic oscillator

\vskip 2mm \tabcolsep 8pt

\centerline{\footnotesize
\begin{tabular}{cccccc}\hline
 Algorithm & FDTD & PSTD & SPSTD & Analytical\\
\hline ${\omega _{000}}$ &9.6606	& 9.9448	& 9.9448	&9.9896 \\
${\omega _{001}}$ &15.9116 	&16.7641	&16.7641	&16.6493 \\
${\omega _{011}}$ &22.1626	&23.2992	&23.2992	&23.3090 \\
${\omega _{111}}$ &28.1295	&30.1185	&29.8343	&29.9687 \\
${\omega _{112}}$ &35.5171	&36.9377	&36.6536	&36.6284 \\
${\omega _{122}}$ &42.0521	&43.7570	&43.4729	&43.2881 \\
${\omega _{222}}$ &49.1556	&51.7128	&50.2922	&49.9479 \\
${\omega _{223}}$ &52.5653	&58.8163	&57.1114	&56.6076 \\
\hline
\end{tabular}}}

\vskip 0.5\baselineskip
\vskip 3mm

\section{Conclusion}
We have developed a SPSTD for solving time-dependent Schr\"{o}dinger equation. On one hand, the scheme has an infinite-order accuracy by using Fourier transforms to represent the spatial derivatives. On the other hand, incorporating the symplectic integrators in the time domain, the scheme demonstrates excellent numerical performances under a long-term simulation. Our numerical results validate significant advantages of the SPSTD scheme in solving the eigenvalue problem of Schr\"{o}dinger equation. The work is fundamentally important for the quantum device simulation.

\ack

This work was supported by the National Natural Science Foundation of China (61301062, 51207041, 61471001,61601166,61701163), the Key Project of Provincial Natural Science Research of University of Anhui Province of China (KJ2015A260).

\end{document}